\documentclass[preprint,aip,apl]{revtex4-2}
\usepackage[utf8]{inputenc}
\DeclareUnicodeCharacter{FB01}{fi}

\usepackage{amsmath}    % need for subequations
\usepackage{graphicx}   % need for figures
\usepackage{verbatim}   % useful for program listings
\usepackage{color}      % use if color is used in text
\usepackage{epstopdf}   % use for side-by-side figures
\usepackage{upgreek}    %for non-italic greek letters (units)
\usepackage[hidelinks]{hyperref}   % use for hypertext links, including those to external documents and URLs
\usepackage{textcomp} %Trademark symbol
\usepackage[separate-uncertainty=true, multi-part-units=single]{siunitx}
\usepackage[capitalise]{cleveref}

\begin{document}

\title{Optical spin-wave detection beyond the diffraction limit}
\author{Juriaan Lucassen}
%\email{j.lucassen@tue.nl,M.J.G.Peeters@tue.nl}
\thanks{These two authors contributed equally. Electronic mail: m.j.g.peeters@tue.nl or r.lavrijsen@tue.nl}
\affiliation{Department of Applied Physics, Eindhoven University of Technology, 5600 MB Eindhoven, the Netherlands}
\author{Mark J.G. Peeters}
\thanks{These two authors contributed equally. Electronic mail: m.j.g.peeters@tue.nl or r.lavrijsen@tue.nl}
%\email{M.J.G.Peeters@tue.nl}
\affiliation{Department of Applied Physics, Eindhoven University of Technology, 5600 MB Eindhoven, the Netherlands}
\author{Casper F. Schippers}
\affiliation{Department of Applied Physics, Eindhoven University of Technology, 5600 MB Eindhoven, the Netherlands}
\author{Rembert A. Duine}
\affiliation{Department of Applied Physics, Eindhoven University of Technology, 5600 MB Eindhoven, the Netherlands}
\affiliation{Institute for Theoretical Physics, Utrecht University, Leuvenlaan 4, 3584 CE Utrecht, the Netherlands}
\author{Henk J.M. Swagten}
\affiliation{Department of Applied Physics, Eindhoven University of Technology, 5600 MB Eindhoven, the Netherlands}
\author{Bert Koopmans}
\affiliation{Department of Applied Physics, Eindhoven University of Technology, 5600 MB Eindhoven, the Netherlands}
\author{Reinoud Lavrijsen}
\affiliation{Department of Applied Physics, Eindhoven University of Technology, 5600 MB Eindhoven, the Netherlands}

\date{\today}
\begin{abstract}
Spin waves are proposed as information carriers for next-generation computing devices because of their low power consumption. Moreover, their wave-like nature allows for novel computing paradigms. Conventional methods to detect spin waves are based either on electrical induction, limiting the downscaling and efficiency complicating eventual implementation, or on light scattering, where the minimum detectable spin-wave wavelength is set by the wavelength of the laser. In this Article we demonstrate magneto-optical detection of spin waves beyond the diffraction limit using a metallic grating that selectively absorbs laser light. Specifically, we demonstrate the detection of propagating spin waves with a wavelength of \SI{700}{nm} using a diffraction-limited laser spot with a size of \SI{10}{\mu m} in \SI{20}{nm} thick Py strips. Additionally, we show that this grating is selective to the wavelength of the spin wave, providing wavevector-selective spin-wave detection. This should open up new avenues towards the integration of the burgeoning fields of photonics and magnonics, and aid in the optical detection of spin waves in the short-wavelength exchange regime for fundamental research.
\end{abstract}

\maketitle
Within magnetism, spin waves are ubiquitous in their presence and applications. Spin waves are fundamental excitations in the magnetization of a material and their behaviour is governed by fundamental magnetic interactions, such as the anisotropy or exchange interaction. Based on this dependence, spin waves are often used to probe these fundamental magnetic interactions.~\cite{PhysRev.110.1295,MAKSYMOV2015253,doppler,Cho2015} In recent years it has been suggested that spin waves can also be used for novel computing methods. Spin-wave propagation occurs without charge transport which allows for computation without Ohmic losses.~\cite{Chumak2015,Chumak_2017} Moreover, spin waves also make excellent candidates for interference based logic devices and non-linear wave computing.~\cite{Chumak2015,Chumak_2017} 

For applications, short-wavelength ($<100$~\si{nm}) spin waves are preferred to reduce the device footprint and increase the group velocity.~\cite{Chumak2017a} The conventional method of exciting and detecting spin waves is based on micron sized microwave antennas through Oersted fields and magnetic induction.~\cite{Chumak2015} Scaling these antennas down such that short wavelength spin waves can be excited and detected ($<100$~\si{nm}), however, is extremely challenging because of the impedance mismatch.  

On the excitation side, several alternatives have been proposed to excite short wavelength spin waves,\footnote{This is by no means a comprehensive list. A more complete overview can be found in Ref.~\citenum{Chumak2015}.} such as spin-transfer torque based methods,~\cite{Demidov2010} grating-like couplers,~\cite{Liu2018,Yu2013} and using the resonance of antiferromagnetically coupled vortex states.~\cite{Wintz2016} However, when it comes to the detection of short-wavelength spin waves progress there are considerably less alternatives. The aforementioned grating couplers can also be used to detect the spin waves. Spin-pumping~\cite{Sandweg2011} and spin-caloritronics~\cite{Schultheiss2012} based techniques do allow for short wavelength detection, but they are not wavelength selective. When using a spectroscopic light-scattering approach, such as Brillouin light scattering, the minimum detectable spin-wave wavelength is determined by the laser wavelength, other optical detection techniques suffer from the diffraction limit and x-ray based techniques have the required resolution but require large scale facilities.~\cite{10.3389/fphy.2015.00035,4773541,Wintz2016} In this Article we therefore demonstrate a grating-based magneto-optical (MO) method to circumvent the diffraction limit when using optical detection of spin waves. It is very much related to the field of near-field optics (see Ref.~\citenum{Rudge2015} and references therein), but simpler to implement than the techniques commonly used to reach super resolution for MO measurements.~\cite{Jersch2010} %Moreover, by using pulsed lasers we also have the time resolution that allows us to measure spin waves.  

\begin{figure}
\centering
\includegraphics[width=\columnwidth]{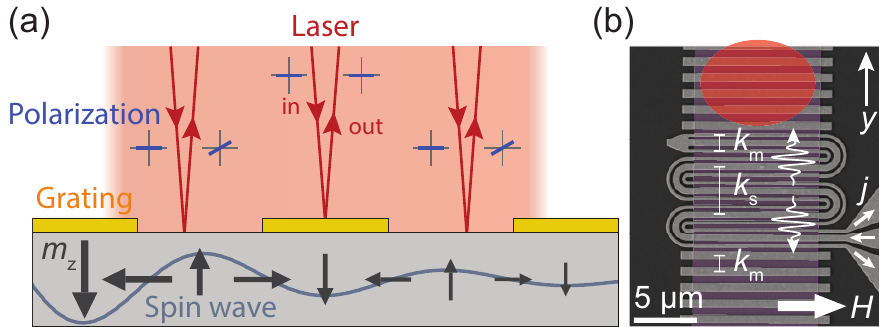}
\caption{\label{fig:figure1}(a) Proposed detection scheme. A laser impinges on a magnetic strip (grey) that contains a spin wave. As this strip is covered with a grating (gold), this modifies the reflection of the incident laser light. The polarization of the incoming laser light that reflects of the magnetic strip (left and right) changes due the magneto-optical Kerr effect in the presence of a spin wave, whilst the polarization of the part that reflects of the grating (middle) remains unaffected, indicated by the rotation of the polarization in the Cartesian coordinate system. (b) SEM micrograph of the fabricated device with $k_\mathrm{m}=5$~\si{\mu m^{-1}}. We drive a microwave current $j$ through the antenna (middle) to generate spin waves with two different wavevectors $k_\mathrm{m,s}$ in the magnetic strip placed underneath. These waves propagate outwards and to measure these spin waves a laser spot (red) is placed on top of the grating (periodicity $k_\mathrm{m}$) following the principle sketched in (a).}
\end{figure}

In~\cref{fig:figure1}a we illustrate the method that we demonstrate in this Article. A metallic grating is placed on top of a magnetic strip with a spin wave. The incident laser light is partially reflected by this grating. However, the light that is transmitted---and is subsequently reflected off the magnetic strip---carries information on the direction of the magnetization through a rotation of its polarization as a result of the magneto-optical Kerr effect (MOKE). Any spin wave whose periodicity matches that of the grating can be detected because the grating acts as a Fourier filter for the magnetization components in the strip. This also allows us to do studies as a function of position by moving the laser spot. By adjusting the relative phase between the probing laser pulses and the excitation current we can achieve phase sensitivity as well. Moreover, this technique will work even when the grating periodicity is smaller than the diffraction limit, although complex scattering effects might need to be taken into account. %Work from the contact-lithography field predicts a near-field resolution of at least $\lambda/20$, with $\lambda$ the wavelength of the light.~\cite{McNab2000} The advantages of this approach are many. In contrast to microwave methods, which suffer from measurement artefacts because of direct antenna-antenna coupling~\cite{Talmelli2018,doi:10.1063/1.5090892} and where spatial dependence studies require new devices to be fabricated, it has several benefits. Direct antenna-antenna coupling is no longer an issue and we can do studies as a function of position because it is easy to probe different areas by moving the laser spot around.  %Moreover, MOKE is very sensitive where precession angles as small as a few m where small precession angles in ultrathin layers are routinely detected. 
%As such, it should aid in investigations of fundamental spin wave properties for small wavelength spin waves. From an application perspective, it is a new step towards the integration of the fields of (nano-) photonics and spintronics.~\cite{Lalieu2019} 
In the remainder of this Article, we start by demonstrating spin-wave excitation through conventional meandered microwave antennas.~\cite{PhysRevB.81.014425} We then move on to actual optical propagating spin wave transmission measurements using the metallic grating. In the last part we focus on an understanding of the measured spectra using a simple analytical model. 

We fabricate devices such as the one shown in~\cref{fig:figure1}b. Here we show a meandering spin-wave antenna located between two gratings. This device allows us to excite spin waves with specific wavevectors belonging to the main periodicities of the spin-wave antenna ($k_\mathrm{m}$ and a small secondary periodicity $k_\mathrm{s}=0.36 \times k_\mathrm{m}$),~\cite{PhysRevB.81.014425} and measure the spin waves by focusing a laser on the grating. The magnetic strip underneath the grating consists of \SI{20}{nm} of Py.
%(\SI{10}{\mu m} wide) underneath the antenna and grating is fabricated using sputtering and an electron beam lithography (EBL) lift-off process. The sputtered stack is //Ta(4)/Py(20)/Al(5) (thicknesses in parentheses in \si{nm}) and was sputter deposited using Ar at $1\times10^{-2}$~\si{mbar} on a Si substrate with a native oxide in a system with a base pressure of $4\times10^{-9}$~\si{mbar}. On top of the magnetic strip, we deposited 40 nm of Al$_2$O$_3$ using atomic layer deposition. Finally, the antennas and grating were created using e-beam evaporation of Ti(10)/Au(100) in a second EBL lift-off process.
Spin waves are measured in the Damon-Eshbach geometry, with the magnetic field and wave vector perpendicular and in the plane. The electrical characteristics of the spin-wave excitation were measured using a vector network analyzer following a procedure described elsewhere.~\cite{doi:10.1063/1.5090892} The optical detection of spin waves was performed using a pulsed laser (\SI{80}{MHz}, pulse length \SI{150}{fs} at a wavelength of \SI{780}{nm}) with a diffraction limited spot size of $\sim 10$~\si{\mu m} measuring the Kerr rotation in combination with continuous wave excitation of the spin waves such that measured magneto-optical (MO) signal is proportional to the spin-wave amplitude at the given phase (the details on the experimental setup and sample fabrication can be found in supplementary note I). Different devices were measured with differing combinations of gratings and spin-wave antennas resonant to spin waves with wavevectors $k=5.0,7.0,9.0$~\si{\mu m^{-1}}, which have wavelengths smaller than our laser spot size and which have long enough attenuation lengths to measure propagating spin waves away from the antenna.
\begin{figure*}
\centering
\includegraphics[width=\textwidth]{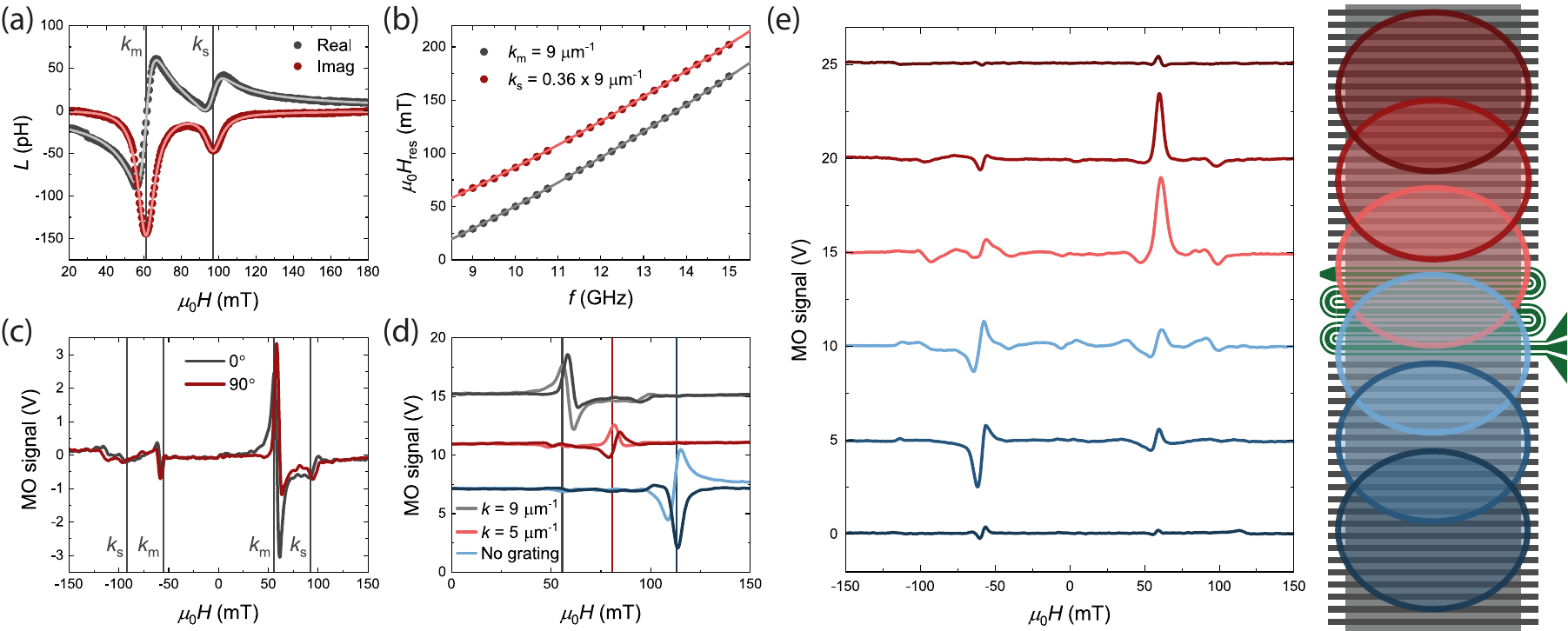}
\caption{\label{fig:figure2}(a) Self induction $L$ for a $k_\mathrm{m}=9$~\si{\mu m^{-1}} antenna including a fit with two (anti)symmetric Lorentzians. We also indicate the resulting resonance fields of the fits for both resonance modes (solid vertical lines). (b) Fitted resonance fields $H_\mathrm{res}$ as a function of frequency $f$ for both resonances [see (a)]. (c) Two phases ($0^{\circ}$ and $90^{\circ}$) of the magneto-optical (MO) signal as a function of magnetic field $H$ for a $k_\mathrm{m}=9$~\si{\mu m^{-1}} device measured at $f=10.24$~\si{GHz} and with the laser spot positioned \SI{12}{\mu m} away from the antenna. The solid vertical lines indicate the resonance fields belonging to the main ($k_\mathrm{m}$) and secondary peak of the spin-wave excitation ($k_\mathrm{s}$). (d) Both phases of the MO signal as a function of magnetic field $H$ for three different gratings optimized for different wavevectors $k$, where no grating corresponds to measurements without a grating present on top of the magnetic strip. The spin-wave antenna generated $k_\mathrm{m}=9$~\si{\mu m} spin waves at \SI{10.24}{GHz}, with the laser spot positioned \SI{12}{\mu m} away from the antenna. The vertical solid lines indicate the resonance field of the spin wave for which the corresponding grating is optimized ($k=0$~\si{\mu m^{-1}} without a grating). Curves are offset for clarity. (e) Several measurements at 1 specific phase for a $k_\mathrm{m}=9$~\si{\mu m^{-1}} device measured at $f=10.24$~\si{GHz} at different positions along the grating, indicated by the sketch on the right. The measurements are spaced~\SI{5}{\mu m} apart and are vertically offset for clarity.}
\end{figure*}

To demonstrate spin-wave excitation, we show the self-induction $L$ of a $k_\mathrm{m}=9$~\si{\mu m^{-1}} antenna in~\cref{fig:figure2}a. In this spectrum two resonances are observed at $60$~\si{mT} and $100$~\si{mT}, which correspond to the excitation of the spin waves with wavelengths equal to the two main periodicities of the antenna ($k_\mathrm{m,s}=9.0, 3.2$~\si{\mu m^{-1}}).~\cite{doi:10.1063/1.5090892,PhysRevB.81.014425} The real and imaginary part of the spectrum are fitted simultaneously with symmetric and anti-symmetric Lorentzian line shapes to extract the resonance fields (solid lines), which are plotted in~\cref{fig:figure1}b as a function of frequency $f$. These resonance fields are then fitted with the dispersion relation for these spin waves (using $M_\mathrm{S}=0.83$~\si{MA.m^{-1}} and $g=2.11$)~\cite{Coey2010,Shaw2013,0022-3719-19-35-014,Kalinikos1981} which results in $M_\mathrm{eff}=0.76 \pm 0.07$~\si{MA.m^{-1}} and thickness $t=16 \pm 5$~\si{nm}, in line with what we expect for Py.~\cite{Shaw2013}

With the verification of spin-wave excitation complete, we move on to the optical measurement of the spin waves. A typical measurement for a device tuned to $k_\mathrm{m}=9$~\si{\mu m^{-1}} is displayed in~\cref{fig:figure2}c, which contains the spin-wave amplitude at two different phases (\ang{0} and \ang{90}) with respect to the microwave excitation source.\footnote{The actual phase difference between the spin-wave amplitude and microwave excitation is offset by an unknown constant value $\phi_0$.} In this measurement there is a resonance at $\sim 55$~\si{mT}, the resonance field of the $k_\mathrm{m}=9$~\si{\mu m^{-1}} spin wave. %spin wave following the dispersion relation of~\cref{fig:figure1}b. 
Furthermore, there is a small resonance at $\sim 100$~\si{mT} which is the result of the large wavelength ($k_\mathrm{s}$) spin waves that are also excited. %and which the grating does allow us to measure (more on this in~\cref{sec:theor_model}). 
A second feature in the spectrum is the amplitude difference between the resonances at positive and negative fields, which is a well-known effect of spin-wave excitation using electrical antennas. It is the result of a difference in spin-wave excitation efficiency because of the magnetic field chirality that either matches or opposes the chirality of the propagating spin wave at positive or negative magnetic fields.~\cite{PhysRevB.77.214411} This is additional confirmation that the grating allows us to detect propagating spin waves with wavelengths of $\sim 700$~\si{nm} ($k_\mathrm{m}=9$~\si{\mu m^{-1}}) using a $10$~\si{\mu m} laser spot.

We can also demonstrate the $k$-selectivity of the grating, which should act as a Fourier filter and be sensitive to only those spin-wave wavelengths that match the grating. This property is illustrated in~\cref{fig:figure2}d, which contains measurements with a $k_\mathrm{m}=9$~\si{\mu m^{-1}} antenna and different gratings designed to detect $k=9,5$~\si{\mu m^{-1}} spin waves and a device without a grating. As a function of the grating $k$-value the resonance field of the measured spin-wave resonance increases, because the gratings are resonant to different spin waves. The predicted resonance fields for the corresponding gratings are indicated in the figure with the solid vertical lines and agree perfectly with the measured resonances. Although the antenna selectively excites spin waves with $k_\mathrm{m}=9$~\si{\mu m^{-1}} there is still a finite excitation efficiency for spin waves with different $k$-vectors. These are measured with gratings that are tuned to different wavelength spin waves and demonstrates the extreme $k$-selectively of the grating. Additionally, for the measurement without a grating there is no resonance at the $k_\mathrm{m}$ peak. This proves that we do indeed need the grating to detect spin waves with wavelengths smaller than the diffraction limit. The large peak visible in the measurement without a grating will be discussed in more detail later in this Article. 

As mentioned in the introduction, an additional benefit of this technique is the possibility to do spatially-dependent measurements of the spin waves. This is demonstrated in~\cref{fig:figure2}e, where several measurements are plotted with the laser spot focused on different positions along the device. There are several interesting aspects here. First, as we move away from the antenna, the propagating spin waves attenuate and resonance amplitudes decrease on a typical length scale of $5-10$~\si{\mu m}, as expected from the spin-wave attenuation length.~\cite{stancil2009spin,Haidar2012} Second, the amplitude asymmetry between the spin waves at positive and negative magnetic field reverses as we move to the other side of the antenna. This is completely in line with the chirality of the excitation mechanism coupled to the type of spin waves that are excited.~\cite{PhysRevB.77.214411} Last, if the laser spot is focused on or near the antenna, the spectrum becomes more complex as extra resonances are present. It is the result of a modified spatial filtering of the spin waves because the laser spot is (partially) filtered by the spin-wave antenna rather than the grating. In supplementary note II we show additional measurements for a $k_\mathrm{m}=5$~\si{\mu m^{-1}} device, which shows similar behavior. To summarize the first part of this Article, we have shown experimentally that the grating technique allows us to measure propagating spin waves with a wavelength of~\SI{700}{nm} using a spot size of $10$~\si{\mu m}. This method enables both wavelength-selective and spatially-dependent measurements.

\begin{figure}
\centering
\includegraphics[width=\columnwidth]{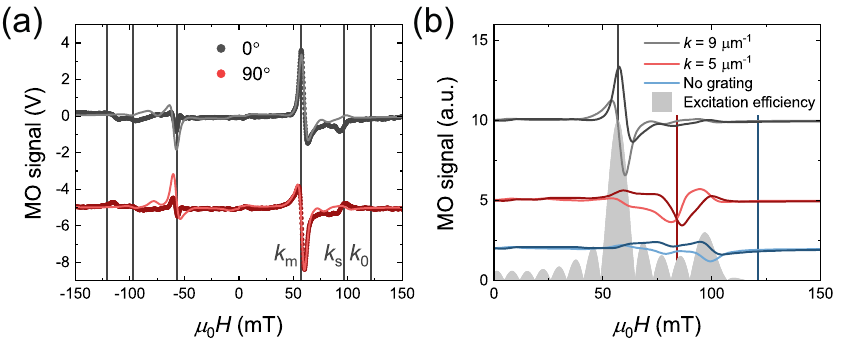}
\caption{\label{fig:figure3}(a) Measurement from~\cref{fig:figure2}c plotted together with the theoretical model, fitted for $H>0$. The solid vertical lines indicate the resonance fields belonging to the $k_\mathrm{m}$, $k_\mathrm{s}$ and uniform FMR peak ($k_\mathrm{0}$). The two phases are offset for clarity. (b) Theoretical calculations for the MO signal as a function of magnetic field for a $k_\mathrm{m}=9$~\si{\mu m^{-1}} antenna and gratings matched to different $k$-values (offset for clarity), using the fit parameters from (a). From top to bottom, the gratings are designed for $k=9.0$ and $5.0$~\si{\mu m^{-1}} as well as a calculation without a grating. The parameters used to calculate these curves are taken from the fit in (a) and the solid vertical lines belong to the resonance field of the $k$-value of the respective grating ($k=0$~\si{\mu m^{-1}} without a grating). The excitation efficiency displays the amplitude of spin waves excited at each field and is determined from the calculated intensity of the microwave excitation field.}
\end{figure}

After the experimental demonstration, we now continue by establishing a more fundamental understanding of the spectra. A simple interpretation of the MO signal is given in~\cref{fig:figure1}a, which suggests the MO signal is determined by a spatial averaging of the filtered laser-spot intensity multiplied by the complete spectrum of excited spin waves. In supplementary note IV we describe how we can calculate the excited spin waves, the filtering of the laser spot and the eventual MO signal using a simple model based on the interpretation given in~\cref{fig:figure1}a. This model is fitted to the measurement of~\cref{fig:figure2}c, as visible in~\cref{fig:figure3}a.\footnote{We use a constrained fit, with the following fit parameters: $\alpha=0.015$, $g=2.1$, $M_\mathrm{s}=0.9 $~\si{MA.m^{-1}}, $M_\mathrm{eff}=0.7$~\si{MA.m^{-1}}, full-width at half maximum of the laser spot equal to~\SI{12}{\mu m}, and laser spot distance from the center of $7$~\si{\mu m}. Parameters that determine the spatial filtering and excited spin waves were fixed and taken from the device geometry. Because of the constrained nature of the fit, and large amount of fit parameters, we do not give uncertainties for the fit.} The model is fitted for positive fields only, but is also plotted at negative magnetic fields.

We find an overall agreement between the measurements and the calculations in~\cref{fig:figure3}a. At positive fields, we find a sharp intense peak around the resonance field (solid line) expected from the $k_\mathrm{m}=9$~\si{\mu m^{-1}} antenna. However, the agreement with the small resonance around the $k_\mathrm{s}$ peak (solid line at $\sim 95$~\si{mT}) is not as good. Although the model does produce a small resonance around this field, the exact positioning, phase and amplitude do not line up with the experimentally measured spectrum. At negative fields the agreement between the measurement and the model is less accurate: there is a significant mismatch in signal amplitude. We do expect a large field asymmetry due to the chirality in the excitation field that either matches or opposes the spin-wave chirality for positive or negative fields. However, the measured amplitude difference between positive and negative fields ($\sim 4$) is very large compared to other electrical and optical measurements ($\sim 2$) which agree with our theoretical calculations.~\cite{Chauleau2014,Haidar2012} %This may be partially explained by the finite penetration depth of the laser light in the metallic layer, combined with the localization of spin waves at the top or bottom of the magnetic layer depending on their chirality. However, with an expected difference in amplitude between the top and bottom of the magnetic layer of $\sim 5\percent$ this is not enough to fully explain the difference between experiments and theoretical calculations.

%Lastly, at negative fields there is another experimental resonance at $\sim -120$\si{mT} (solid line) that is not present in the calculations or at positive fields.

The model can also reproduce the $k$-selectivity of the grating qualitatively as illustrated in~\cref{fig:figure3}b, where the calculated theoretical curves for several different gratings parameters are plotted. As the grating $k$-value goes down, the main resonance moves up to higher magnetic fields. Moreover, the intensity decreases as only the top MO curve has a grating matched to the main excited spin wave. Additionally, we find that the position of these resonances for the $k=5$ and $k=9$~\si{\mu m^{-1}} grating match very closely to what is expected based on the dispersion relationship (solid vertical lines) and what is observed experimentally (\cref{fig:figure2}d). We understand these spectra by realizing that the antenna excites a wide-range of different-wavector spin waves, which we visualize as the excitation efficiency with the grey shaded area in~\cref{fig:figure3}b. Even though the grating is not matched to the spin-wave antenna, there is still a small amount of spin waves that are excited which have wavelengths matched to the grating. 

Yet, the spectrum for the case without a grating differs from the experimental situation (\cref{fig:figure2}d). In the experiments, there is a resonance at approximately the uniform FMR mode. The calculations, however, show that this resonance should not be excited by the spin-wave antenna. Something similar is also observed around~\SI{-120}{mT} ($k_\mathrm{0}$) in the measurements where a grating is present, as shown in~\cref{fig:figure3}a. As electrical spin-wave excitation should not occur at these magnetic fields, we cannot explain the presence of these peaks. In supplementary note V we provide additional details on these peaks from which we conclude that it is not a spin-wave mode that is excited by the laser. %Moreover, the fact that the resonance is not present with a grating present on the strip suggests that the absence of the grating itself is essential in exciting this FMR-like resonance.

Although we do not know the origin of the uniform FMR like peak in the measured spectra, we do not believe it affects the main conclusions of this Article. The $k$-selectivity of the grating, the exact location of the resonance fields and the frequency dependence of the resonances (not shown) all suggest that the grating can be used to detect spin waves with a wavelength smaller than the laser spot. Finally, in supplementary note II we also show calculations for a position sweep and compare them with the experimental results, showing that the model is also able to qualitatively explain this behavior.

%This is the ultimate limit that can be reached using blue light ($\lambda=400$~\si{nm})~\cite{Chauleau2014}, but we have not pushed the ultimate limit of this technique.  
Thus, overall, the theoretical model can qualitatively explain the most relevant data. 
%Yet, as detailed in the previous paragraphs, there are some features in the experiments that we do not find in the theoretical curves. For one, the measured amplitude asymmetry between positive and negative fields is much larger than expected based on the excitation efficiency difference. Second, the $k_\mathrm{s}$ resonance is not well-described by the calculations. 
Yet, the inability of the model to fully explain the data suggests that the simple approach we advocate here for a qualitative understanding of the measured spectrum is incomplete. There are some indications that this might be related to the complex optical behaviour of the grating. As we demonstrate in supplementary note III, light polarized parallel to the grating produces a finite MO signal, whilst for light polarized perpendicular to the grating lines there is no MO contrast. Moreover, in this same supplementary note we also demonstrate that the measured MO contrast depends on the $k$ value of the grating, with a maximum signal around $k=9$~\si{\mu m^{-1}}. Last, we also show that if we place the grating directly on top of the strip, rather than on top of the insulating layer, the measured signal increases significantly, suggesting the extra optical path length through the insulating layer leads to spreading out of the light which reduces the $k$-sensitivity of the grating. To fully understand this behavior, more detailed theoretical and experimental work should be performed to investigate the true influence of the grating. Near-field effects such as plasmonic resononances need to be carefully considered to form a complete picture of the behaviour of the grating.~\cite{McNab2000,RevModPhys.82.729} In this case specifically, the effect of the grating on the polarization of the light needs to be addressed. Last, we have made no attempt to optimize the thickness of the grating, although one can image that this can also impact its behavior.

Nonetheless, theoretical work from contact lithography suggests that the resolution in the near-field can be as high as $\lambda/20$,~\cite{McNab2000} limited only by the distance between the grating and the magnetic strip, and required MO contrast to produce a measurable signal. This demonstrates that our technique has enormous potential for the detection of spin waves with wavelengths as small as several tens of \si{nm}. One can then imagine a wealth of applications for this method that range from fundamental investigations into spin-wave behavior to the practical investigations into magnonic devices. For example, one can investigate $k$-dependent behaviour as a function of position using an optical technique with wavelengths that are usually only accessible using x-ray based techniques.~\cite{4773541,Wintz2016} More on the practical side, it allows one to, for example, measure the iDMI in films thinner than reachable using all-electrical propagating spin-wave spectroscopy.~\cite{Lucassen2020} As ultrathin films are routinely measured with MOKE, we expect the layer thickness to be less of an issue for the optical technique. %We are currently in the process of trying to measure the iDMI in \SI{5}{nm} thick Pt/Co/Ir films using this technique. 
Additionally, the inverse of the process demonstrated here should also be possible: using a grating to $k$-selectively excite spin waves with an ultra-fast laser pulse.~\cite{physRevLett.110.097201,PhysRevLett.88.227201} This brings nano-scale all-optical propagating spin-wave spectroscopy one step closer to realization.~\cite{Hashimoto2017} %Last, from an application perspective it is a new step towards the integration of the fields of (nano-) photonics and spintronics.

Concluding, in this Article we have demonstrated an optical technique that can measure propagating spin waves beyond the diffraction limit using a metallic grating. This technique works down to wavelengths of at least~\SI{700}{nm} ($k_\mathrm{m}=9$~\si{\mu m^{-1}}) using a diffraction-limited laser spot size of~\SI{10}{\mu m}. Additionally, we could also detect spin waves with extreme $k$-selectivity and investigate the propagation of spin waves as a function of position. This opens up a route for the integration of photonics and magnonics,~\cite{Lalieu2019} and brings us one step closer to optical detection of spin waves in the exchange-wave regime. 

\begin{acknowledgments}
This work is part of the research programme of the Foundation for Fundamental Research on Matter (FOM), which is part of the Netherlands Organisation for Scientific Research (NWO). We also thank L.G.T. van de Coevering from March Microwave Systems B.V. for lending us microwave equipment.

\section*{Author declarations}
\subsection*{Conflict of Interest}
The authors have no conflicts to disclose.

\subsection*{Author Contributions}
\textbf{J. Lucassen:} Investigation (equal), writing -- original draft (lead). \textbf{M.J.G. Peeters:} Investigation (equal), writing -- original draft (supporting). \textbf{C.F. Schippers:} Methodology, writing -- review and editing (equal). \textbf{R.A. Duine:} Supervision (equal), writing -- review and editing (equal). \textbf{H.J.M. Swagten:} Supervision (equal), writing -- review and editing (equal). \textbf{B. Koopmans:} Conceptualization, supervision (equal), writing -- review and editing (equal). \textbf{R. Lavrijsen:} Supervision (equal), writing -- review and editing (equal). 

\section*{Data Availability Statement}
The data that support the findings of this study are available from the corresponding authors upon reasonable request.

\end{acknowledgments}
\bibliography{references}
%%\bibliography{library,Addition}
\end{document}

%\title{test}

\title{Supplementary Material: Optical spin-wave detection beyond the diffraction limit}
\author{Juriaan Lucassen}
%\email{j.lucassen@tue.nl,M.J.G.Peeters@tue.nl}
\thanks{These two authors contributed equally. Electronic mail: j.lucassen@tue.nl or M.J.G.Peeters@tue.nl.}
\affiliation{Department of Applied Physics, Eindhoven University of Technology, 5600 MB Eindhoven, the Netherlands}
\author{Mark J.G. Peeters}
\thanks{These two authors contributed equally. Electronic mail: j.lucassen@tue.nl or M.J.G.Peeters@tue.nl.}
%\email{M.J.G.Peeters@tue.nl}
\affiliation{Department of Applied Physics, Eindhoven University of Technology, 5600 MB Eindhoven, the Netherlands}
\author{Casper F. Schippers}
\affiliation{Department of Applied Physics, Eindhoven University of Technology, 5600 MB Eindhoven, the Netherlands}
\author{Rembert A. Duine}
\affiliation{Department of Applied Physics, Eindhoven University of Technology, 5600 MB Eindhoven, the Netherlands}
\affiliation{Institute for Theoretical Physics, Utrecht University, Leuvenlaan 4, 3584 CE Utrecht, the Netherlands}
\author{Henk J.M. Swagten}
\affiliation{Department of Applied Physics, Eindhoven University of Technology, 5600 MB Eindhoven, the Netherlands}
\author{Bert Koopmans}
\affiliation{Department of Applied Physics, Eindhoven University of Technology, 5600 MB Eindhoven, the Netherlands}
\author{Reinoud Lavrijsen}
\affiliation{Department of Applied Physics, Eindhoven University of Technology, 5600 MB Eindhoven, the Netherlands}

\date{\today}

\maketitle
\section{Experimental details}
\label{sec:opt_det}
\begin{figure}[!tb]
\centering
\includegraphics[width=1.0\textwidth]{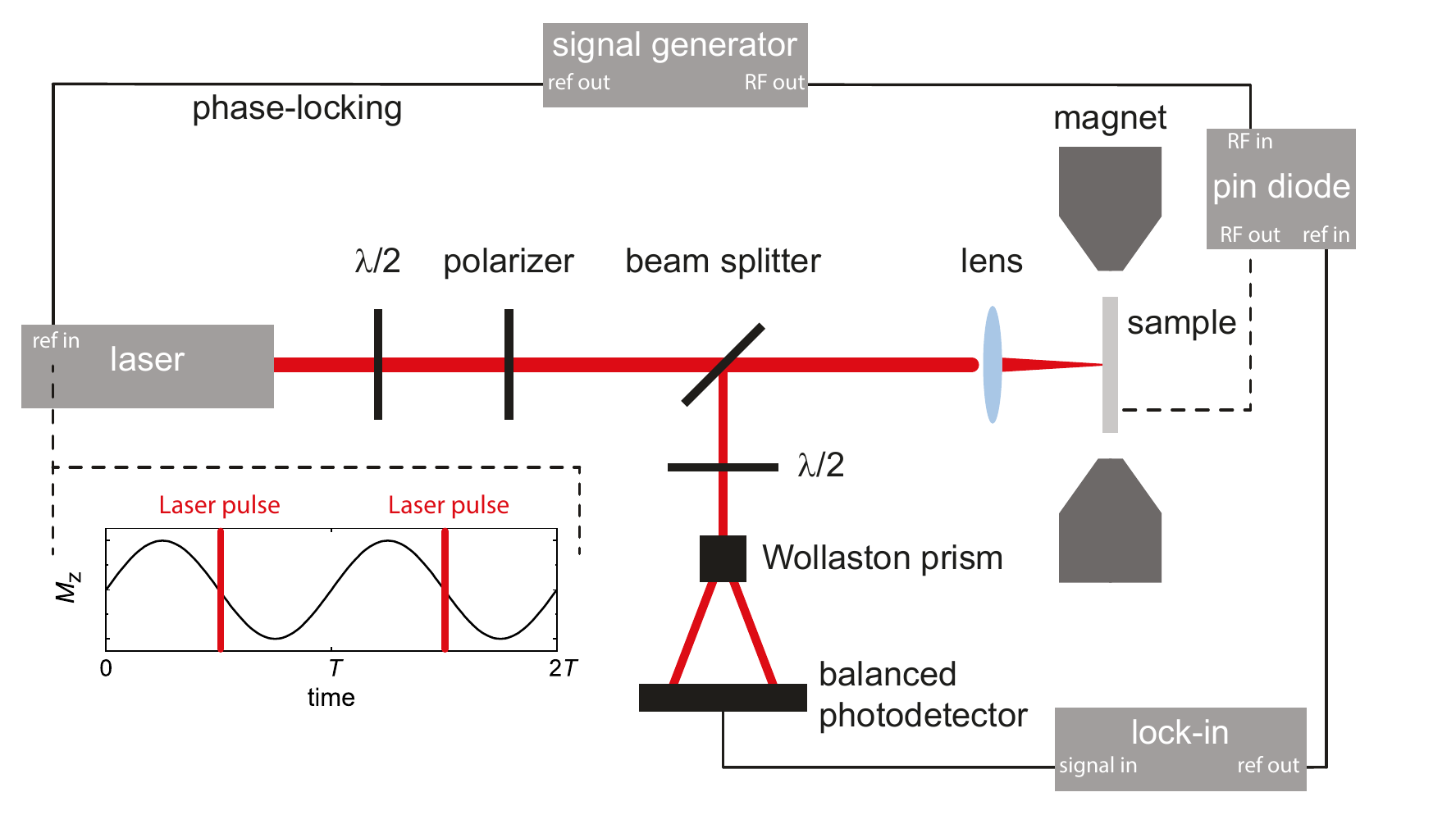}
\caption{\label{fig:setup} Sketch of the experimental setup. A pulsed laser is focused onto the sample in which we excite spin waves using a RF signal generator. This signal from the RF signal generator is phase-locked to the laser, such that the incoming laser pulse is always probing the same phase of the oscillating out-of-plane component $M_\mathrm{z}$ of the spin wave (see bottom left). We also modulate the power of the RF output using a pin-diode. The reflected laser signal is separated using a beam-splitter and focused onto a balanced photodector, where we detect the change in the polarization of the light. The magneto-optical (MO) signal is then read out using a lock-in detector locked to the modulation frequency of the pin-diode.}
\end{figure}

To measure the MOKE signal from a spin wave, we utilized the experimental setup that is illustrated in~\cref{fig:setup}. The setup can be divided into three main components. We have the optical measurement of the spin waves using MOKE, electrical RF excitation of the spin waves using the signal generator, and the phase locking of the laser pulses to the electrical excitation. In the following, we will discuss each of them individually to understand the setup and the performed measurements.

For the optical detection of the spin waves, we need a pulsed laser. As the magnetization of the spin wave oscillates in time, a time-averaging of the MOKE signal from the spin wave will result in a vanishing signal. We therefore use a Ti:Sapphire pulsed laser with an approximate pulse length of~\SI{150}{fs}, a wavelength of \SI{780}{nm}, and a repetition rate of~\SI{80}{MHz} that samples the spin wave at a predefined phase (this will be discussed in more detail later). The laser spot is focused down to a diffraction-limited spot of about $10$~\si{\mu m} using a lens with an NA of $0.38$.\footnote{We do not completely fill the objective. When doing so, the spot size will be reduced even further.} To be sensitive to only the out-of-plane component of the magnetization, we need to come in at perpendicular incidence. For that reason, a beam splitter is used. The first half-wave plate ($\lambda/2$) and polarizer are used to tune both the power of the linearly polarized output of the laser, as well as set the polarization of the probing laser pulse on the sample. A change in polarization due to MOKE is measured using a Wollaston prism in combination with a balanced photodetector. The prism separates the incoming light into two polarized light beams with orthogonal polarization, and the photodetector produced a signal that is proportional to the difference in intensity between these two beams. Another half-wave plate is used to balance the photo-detector by rotating the polarization of the light such that the intensity of the two polarized beams is equal and the output of the photodetector is negligible without the presence of spin waves. Any small change in the polarization in the light due to a MOKE signal will then lead to a difference between the outputs of the Wollaston prism, and to a signal on the detector that is proportional to change in polarization of the light. Thus, the measured signal on the photodetector is proportional to the amount of spin waves that are excited.

To excite the spin waves, we used an RF signal generator (Anritsu MG3692C) with an output power of \SI{15}{dBm}. The RF generator is connected to the spin-wave antenna using microwave probes. We modulated the output of this generator using an RF pin diode ($100$~\si{kHz}) to increase signal-to-noise ratio as this allowed us to lock onto this modulation frequency with the lock-in when measuring the signal from the balanced photodetector.

An important element of the setup, which we have so far neglected, and which we illustrate at the bottom left in~\cref{fig:setup} is the phase-locking of the incoming laser-pulse to the RF excitation. Pulsed lasers are needed to avoid a vanishing time-averaged signal and for that same reason, we need to ensure that every incoming laser-pulse probes the exact same phase of the spin wave. If this is not the case, the time-averaged signal we measure also vanishes. The details on the phase stabilization between the laser and RF signal generator are not shown in the figure. It was achieved by monitoring the relative phase between the two on a \SI{6}{GHz} oscilloscope and adjusting the phase of the laser accordingly. This was also used to vary the relative phase between the excitation current and probing laser pulse such that we could probe at different times within a spin wave. To be able to measure high-frequency RF signals on the scope, we used a~\SI{-30}{dB} directional coupler in combination with a by-$8$ frequency divider to redirect a small fraction of the RF signal into the scope with a detectable frequency. 

The devices such as the one shown in Fig. 1b of the main paper are fabricated as follows. The magnetic strip (\SI{10}{\mu m} wide) underneath the antenna and grating is fabricated using sputtering and an electron beam lithography (EBL) lift-off process. The sputtered stack is //Ta(4)/Py(20)/Al(5) (thicknesses in parentheses in \si{nm}) and was sputter deposited using Ar at $1\times10^{-2}$~\si{mbar} on a Si substrate with a native oxide in a system with a base pressure of $4\times10^{-9}$~\si{mbar}. On top of the magnetic strip, we deposited \SI{40}{nm} of Al$_2$O$_3$ using atomic layer deposition. Finally, the antennas and grating were created using e-beam evaporation of Ti($10$)/Au($100$) in a second EBL lift-off process. 
\section{Additional measurements}
\label{sec:add_meas}
In the main paper we presented measurements for a $k_\mathrm{m}=9$~\si{\mu m^{-1}} antenna at~\SI{10.24}{GHz}. In this supplementary we show an additional position sweep for a $k_\mathrm{m}=5$~\si{\mu m^{-1}} device, including the results of the theoretical model.

\begin{figure}
\centering
\includegraphics[width=\textwidth]{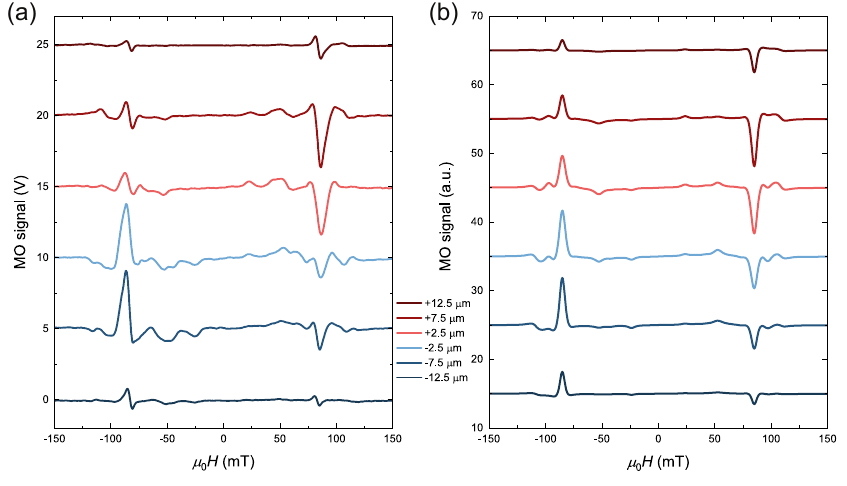}
\caption{\label{fig:extra_meas}  Results of measurements (a) and simulations (b) at 1 specific phase for a $k_\mathrm{m}=5$~\si{\mu m^{-1}} device at $f=10.24$~\si{GHz} for different laser spot positions along the grating (given by the $y$-position indicated in Fig.~1a of the main paper, with $y=0$~\si{\mu m} the middle of the antenna). The curves are spaced~\SI{5}{\mu m} apart and are vertically offset for clarity.}
\end{figure}
The results of the measurements are displayed in~\cref{fig:extra_meas}a, where the main resonance peak has shifted to higher fields ($\approx 85$~\si{mT}) in agreement with the shorter wavelength of the spin wave. Similar to the $k_\mathrm{m}=9$~\si{\mu m^{-1}} results, we observe that as we move away from the antenna, the spin waves attenuate with length scale of $5-10$~\si{\mu m}. We additionally also find that the amplitude asymmetry between the spin waves at positive and negative magnetic field reverses as we move to the other side of the antenna.  The theoretically calculated curves, using the fit-parameters from the fit in Fig. 3a of the main paper, are plotted in~\cref{fig:extra_meas}b. There is an overall qualitative agreement between the measurements and calculations, although small individual details do vary between measurements and calculations.
\section{Optical behavior of grating}
\label{sec:opt_opticalcomplex}
\begin{figure}[!bt]
\centering
\includegraphics[width=\textwidth]{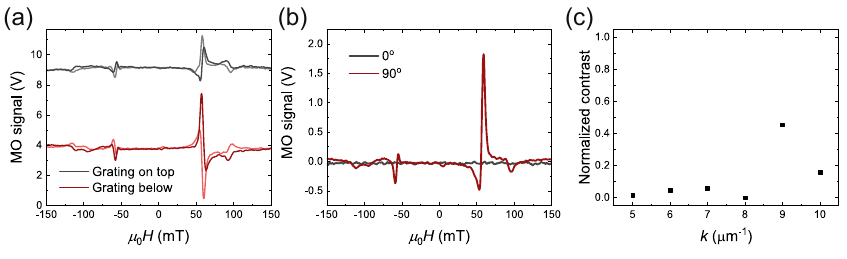}
\caption{\label{fig:grating_comb}(a) Both phases of the MO measurements when the grating is on top of the insulating layer and when the grating is positioned below the insulating layer with $k_\mathrm{m}=9$~\si{\mu m^{-1}} at~\SI{10.24}{GHz}. (b) MO optical trace for a situation with the incoming light polarization at~\ang{0} (perpendicular to grating lines) or at~\ang{90} (parallel to grating lines) with $k_\mathrm{m}=9$~\si{\mu m^{-1}} at~\SI{10.24}{GHz}. (c) Magnetic contrast as a function of grating periodicity $k$. Step size of hysteresis loop measured on out-of-plane magnetized Pt/Co/Pt sample measured through gratings normalized to the step-size measured without a grating.}
\end{figure}

In this section we show three elements which indicate that the optical behavior of the grating is more complex than suggested in Fig. 1a of the main paper. First, in~\cref{fig:grating_comb}a we show that the exact position of the grating is relevant for the amplitude of the magneto-optical (MO) signal. When the grating is placed below the insulating layer, the signal is enhanced with respect to the the situation where the grating is placed on top of the insulating layer. This might be the result of the extra optical path length through the insulating layer leading to additional diffraction and reducing the $k$-sensitivity of the grating.~\cite{McNab2000}

Second, in~\cref{fig:grating_comb}b the dependence of the MO signal on the polarization of the incoming light is displayed. When the polarization is parallel to the grating (\ang{90}) there is magnetic contrast, which has vanished upon rotating the polarization (\ang{0}). This behavior is contrary to one what would expect when one considers the grating as a wire polarizer.~\cite{RevModPhys.82.729,McNab2000}  The transmission through the grating should then be maximum for a polarization of \ang{0}. In contrast, the part of the reflected light with the induced Kerr rotation from the spin waves will experience minimum transmission because it has a rotated polarization. The opposite is true for incoming light with a polarization of \ang{90}. In this case, the incoming light will be minimally transmitted, but the transmission of reflected light with its Kerr rotation will be large. Because of this, the reflected light with the rotated polarization due to the Kerr effect will be equal for both cases and we would not expect a dependence on the polarization of the incoming light.

Third, \cref{fig:grating_comb}c shows that the MO contrast varies with the grating wavevector $k$ in a non-trivial manner with a maximum around $k=9$~\si{\mu m^{-1}}. For these measurements, we placed gratings of different $k$-values on a full sheet Pt/Co/Pt stack. The contrast was determined by extracting the step size from out-of-plane hysteresis loops measured through the grating and then normalized to a step size measured on an area without a grating. We do not know why the MO contrast depends on $k$, but it is not in accordance with the simple picture sketched in~\cref{fig:grating_comb}a in which it should not depend on $k$. These three different elements of the grating demonstrated here suggest a more complete picture needs to be formed through detailed simulations of the optical behaviour of the grating. We expect that near-field effects such as plasmonics will play a large role.~\cite{McNab2000,RevModPhys.82.729}
\section{Model details}
\label{sec:mod_details}
The model we use to qualitatively understand the data combines two different elements:
\begin{enumerate}
\item The excitation of the spin waves from the microwave antenna.
\item The selective blocking of the laser light from the grating and antenna.
\end{enumerate}
If we combine these effects, the selectively blocked laser light reflects of the magnetic strip with the excited spin waves, leading to an average Kerr rotation proportional to the product of the laser light and the magnetization of the spin waves.

We first calculate the excited spin waves, and to do this we realize that (in $k$-space)
\begin{equation}
\label{eq:susp}
\mathbf{m}_{k}=\chi_{k} \mathbf{h}_{k},
\end{equation}
with $\mathbf{m}_{k}$ the (complex) amplitude of the spin wave with wavevector $k$ in the $y$-direction, $\mathbf{h}_{k}$ the magnetic field and $\chi_{k}$ the Polder spin-wave susceptibility tensor.~\cite{doi:10.1080/14786444908561215} To calculate the total spin-wave spectrum $\mathbf{m}_{k}$, we simply calculate the Oersted fields from the antenna, Fourier transform these, and plug them into~\cref{eq:susp}. There is an inherent assumption here as we do not solve Maxwell's equations fully to calculate $\mathbf{m}_k$. Instead, we assume that $h_\mathrm{k}$ is determined by the current running through the antenna. This is not true because of the presence of the grating and the magnetic strip, which modifies the magnetic field distribution, although calculations demonstrate that it does not modify the situation drastically.~\cite{PhysRevB.81.014425}

The selective blocking of the laser light leads to a magneto-optical signal $\mathrm{MO}$ in perpendicular incidence proportional to the out-of-plane magnetization $m_\mathrm{z}$ averaged over the laser spot $L$:
\begin{equation}
\mathrm{MO} \propto  \int \mathrm{d}y~m_\mathrm{z} L(y),
\end{equation}
where we explicitly introduced the spatial profile of the laser spot through its dependence on $y$ (see Fig. 1b of main paper). This laser spot $L$ has a Gaussian profile modified by the absorption of both the grating, and if its partly on top of the antenna, the antenna as well. For this equation to be valid, several assumptions must be made:
\begin{enumerate}
\item We assume that the thickness of the grating is much thicker than the extinction depth of the light such that the blocking mechanism sketched in Fig. 1a of the main paper is valid, and
\item we completely disregard any near-field effects and plasmonics.
\end{enumerate}
The last assumption, although critical for us to be able to construct an easy model, is an oversimplification as we have demonstrated in~\cref{sec:opt_opticalcomplex}.

Combining both equations produces
\begin{equation}
\mathrm{MO} \propto  \int \mathrm{d}y~\mathcal{F}^{-1}(m_\mathrm{k,z}) L(y),
\end{equation}
with $\mathcal{F}^{-1}$ the inverse Fourier transform. This is what is calculated in this Article, where we use the relevant magnetic parameters in the susceptibility tensor $\chi$, the antenna (and device) parameters to calculate $\mathbf{h}$, and put in the information about the laser spot and the grating/antenna in $L$.
 %Second, and perhaps more important, we greatly simplify the effect of the grating. %As detailed in~\cref{sec:opt_opticalcomplex} the grating has a non-trivial effect on the polarization and transmission of the incoming laser that is not included in the model.

%We assume that the grating does not effect the linear dependence of the Kerr rotation on $m_\mathrm{z}$, an assumption greatly challenged by the strong polarization dependence of the grating (see sec.~\cref{sec:opt_opticalcomplex} for experimental confirmation and Ref.~\citenum{McNab:00} for theoretical confirmation). Moreover, we assume that we can decouple the behaviour of the grating and its interaction with the spin waves. I.e. there is an absorption through the grating, and then a Kerr rotation through the interaction with the spin waves upon reflection. Due to the complex nature of the grating, this decoupled nature might be too simple an approximation and a more complex picture might be needed, where we for example look at the Reflection coefficients in $k$ space to account for different sensitivities to different $k$ spin waves. However, this is beyond the scope of this paper.
\section{Uniform mode}
\label{sec:uniform}
In this section we give some details on the FMR-like (hereafter referred to as uniform) resonance peak we find in both the measurements with a grating (the small $k_\mathrm{0}$ peak in Fig. 3a of the main paper), and the large peak in the measurement without the grating (Fig. 2d of the main paper). As we indicate in Fig. 3b of the main paper, the uniform FMR mode should not be excited by the spin-wave antenna. This also corresponds with the data presented in Fig. 2a of the main paper, where we see no resonances excited near the FMR-field. The presence of these peaks is therefore surprising. Although we do not know what the origin is of these resonances, we point out some peculiarities of the behavior of these resonances:   %Lastly, we note that it is as of yet unclear as the whether the uniform resonance in the measurements with and without the grating is of the same origin. 
%Although we do not know what the origin is of these resonances, we point out some peculiarities of the behavior of this resonance.
%There are couple of things we note: 
\begin{enumerate}
\item Because we are able to measure the uniform resonance with a~\SI{10}{\mu m} laser spot, and because its resonance field suggests it is a uniform mode, we conclude that this is a short-wavelength resonance. This means that we should also be able to measure it with the grating, which is as selective for the $k$-value for which it was designed, as for the uniform FMR mode. However, as we show in Fig. 3a of the main paper, it is only very weakly present when there is a grating on the strip. This can be explained if there is a huge difference in the detection efficiency between the situation with and without a grating. As we demonstrate in~\cref{sec:opt_opticalcomplex} this is the case, but for the observed detection efficiency we would still expect a larger amplitude of the uniform peak when there is a grating present than what we see in the experiments (for example in Fig. 2d of the main paper). 
%\item We conclude that this resonance is not the result of parametric pumping. This will yield resonances at half the frequency, and thus at lower magnetic fields rather than higher magnetic fields (see for example Ref.~\citenum{PhysRevB.89.104422} and references therein).
%\item It is not heat related as it does not scale linearly with the RF power.
\item The uniform FMR mode is not laser-induced. Turning off the RF power removed the resonance.
\item In the situation with a grating (Fig. 2e of the main paper), we find that below the antenna, the amplitude of the uniform mode is higher for positive fields, and above the antenna, higher at negative fields. This is in contrast to the amplitude asymmetry for the $k_\mathrm{m}$ and $k_\mathrm{s}$ mode (Fig. 2e of the main paper). In contrast, without a grating the amplitude asymmetry for the uniform mode is similar to the $k_\mathrm{m}$ and $k_\mathrm{s}$ mode (not shown). This is surprising for three reasons:
\begin{enumerate}
\item The presence of the small uniform peak far away from the excitation antenna suggests it is a propagating wave, which should not exist at these fields according to the dispersion relation.
\item If it is indeed a uniform mode, there should be no difference in amplitude between the bottom and top of the antenna.
\item If the uniform mode for the situation with and without the grating is excited by the same mechanism, why is the magnetic field amplitude asymmetry reversed between both cases?
\end{enumerate}
\end{enumerate}
\bibliography{../references}